%Paper: hep-th/9305144
%From: stroganov@mx.decnet.ihep.su
%Date: Wed, 26 May 1993 12:12:07 +0400

\magnification=\magstep2

\def\det{{\rm\bf det}}
\def \w {\omega}
\def \s {\sigma}
\def \R {{\cal R}}
\pageno=0
{\nopagenumbers
\line{\hfil May, 1993}
\vskip 6 true cm
\centerline {\bf Generalized Yang-Baxter Equation}
\vskip 2 true cm
\centerline {R.M. Kashaev and Yu.G. Stroganov
\footnote*{Institute for High Energy Physics,
Protvino, Moscow Region 142284, Russia}}
\vskip 1 true cm
\centerline {\it St. Petersburg Nuclear Physics Institute,}
\centerline {\it Gatchina, St. Petersburg 188350, Russia}
\vskip 2 true cm
\line{\bf Abstract \hfil}

A generalization of the Yang-Baxter equation is proposed. It
enables to construct integrable two-dimensional lattice models
with commuting two-layer transfer matrices, while single-layer
ones are not necessarily commutative. Explicit solutions to the
generalized equations are found. They are related with Botzmann
weights of the $sl(3)$ chiral Potts model.
\vfil
\eject}

\beginsection{1. Introduction}

As is known the Yang-Baxter equation (YBE) ensures commutativity
of one-layer transfer matrices (TM's) in two-dimensional vertex lattice
models [1]. Obviously, two and more-layer TM's commute among themselves
in this case as well. Is it possible to have commuting family of e. g.
two-layer TM's while
 one-layer ones being non-commutative? Clearly, a positive
answer to this question would provide us with a wider class of solvable
lattice models.
In this paper we propose a generalized YBE, which realizes such a
 possibility
(provided the ''cross'' matrices of Boltzmann weights are non-degenerate).
The generalized equation has the following form:
$$
\eqalign{
&\sum_{k_1,k_2,k_3}S(p,q)_{i_1,i_2}^{k_1,k_2}
\overline S(p,r)_{k_1,i_3}^{j_1,k_3}
S(q,r)_{k_2,k_3}^{j_2,j_3}\cr
=&R_{p,q,r}\sum_{k_1,k_2,k_3}
\overline S(q,r)_{i_2,i_3}^{k_2,k_3}
S(p,r)_{i_1,k_3}^{k_1,j_3}
\overline S(p,q)_{k_1,k_2}^{j_1,j_2},\cr}           \eqno(1.1a)
$$
where all indices run over $N\ge2$ distinct values $0,1,\ldots,N-1$,
 ``rapidities'' $p,q,r$ represent some continuous variables, and
$S(p,q)^{k,l}_{i,j}$, $\overline S(p,q)^{k,l}_{i,j}$, $R_{p,q,r}$ are
 functions of their arguments. In the standard way one can write eqs. ($1.1a$)
in a matrix form:
$$
S_{12}(p,q)\overline S_{13}(p,r)S_{23}(q,r)=
R_{p,q,r}\overline S_{23}(q,r)S_{13}(p,r)\overline S_{12}(p,q), \eqno(1.1b)
$$
where, for example, $N^3$-by-$N^3$ matrix $S_{12}(p,q)$ acts on basis
ket-vectors as follows:
$$
S_{12}(p,q)|i,j,k\rangle=\sum_{i',j'}|i',j',k\rangle
S(p,q)^{i,j}_{i',j'},                                \eqno(1.2)
$$
the other matrices being defined similarly.
If we put in $(1.1b)$ $R_{p,q,r}=1$ and $\overline S=S$, then we recognize
the well known ``vertex'' YBE [2], [1]. In this sense eqs. (1.1) are a
generalization of the latter.
 Note, the function
$R_{p,q,r}$, up to some root of unity, can be absorbed by
multiplicative redefinition of $S$, $\overline S$ matrices (of course
if they are non-degenerate). Indeed, calculating determinants of both sides
of $(1.1b)$, we see that
$$
(R_{p,q,r})^{N^3}=f_{p,q}f_{q,r}/f_{p,r}                     \eqno(1.3)
$$
 for some $f$'s. Renormalizing now $\overline S(p,q)\to\overline
S(p,q)/(f_{p,q})^{1/N^3}$ we eliminate the $R$ function from (1.1) up to an
$N^3$-th root of unity.
Nevertheless, we will keep $R$-factor in an explicit form to avoid possible
problems with branching.

Note that
eqs. (1.1) resemble the star-triangle relations [1], which can be
written as
$$
\eqalign{
W_{p,q}(m,k)&\overline W_{p,r}(k,l)W_{q,r}(m,l)\cr
           =&R_{p,q,r}\sum_n\overline W_{q,r}(k,n)W_{p,r}(m,n)
            \overline W_{p,q}(n,l),\cr}                        \eqno(1.4)
$$
where $W$'s and $\overline W$'s correspond to $S$'s and $\overline S$'s
respectively, the difference being in number of discrete arguments as well
as summations.

We should also notice a resemblance of eqs. (1.1) with ``twisted'' YBE
of refs.[3], [4]\footnote*{We thank L.D. Faddeev, A.Yu. Alekseev,
E.K. Sklyanin for pointing out to these
references and discussion}.
In the ``twisted'' YBE from these papers matrices, corresponding
to our $\overline S$'s, act nontrivially in all three sub-spaces of the
tensor product space. Particular solutions from [3], however, can not
be specialized to non-trivial solutions of (1.1). If one demands the analogs
of our $\overline S$'s to be trivial in third sub-space, then the ``twisted''
YBE reduces to the usual one. Most general ``twisted''
YBE of [4] appeared in the context of quasi-Hopf algebras. From this point
of view, eqs. (1.1) seem to correspond to a particular quasi-Hopf algebra.
Unfortunately, the latter is far from to be clear for us, so we will not
discuss this point anymore in this paper.

In Section 2, assuming non-degenerateness conditions in the ``cross
channel'' for the matrices $S$, $\overline S$, satisfying (1.1),
we show that $N^2$-by-$N^2$
box-matrices, constructed in terms of matrices $S$, satisfy the usual YBE.
As a consequence, the corresponding TM's, being in fact two-layer ones,
commute among themselves. In Section 3 explicit solutions to (1.1)
are presented. In Section 4 the results obtained are summarized with
some discussion. In Appendix A the functions of Sect. 3 are proved
to satisfy eqs. (1.1). Appendix B contains an explanation of why in the case
of even number of local states our solutions are degenerate in the ``cross
channel''.

\beginsection{2. ``Box'' Construction}

In this section we assume the non-degenerateness condition
in the ``cross channel'':
$$
\det S^{t_2}(p,q)\ne0,\quad \det\overline
S^{t_2}(p,q)\ne0,                                       \eqno(2.1)
$$
where $S^{t_2}$, $\overline S^{t_2}$
are matrices $S$, $\overline S$, transposed in the second space. Introduce
inverse ``cross''matrices $S'$, $\overline S'$, satisfying
$$
\sum_{j,k}S(p,q)_{i,j}^{k,l}S'(q,p)_{m,k}^{j,n}=
\delta_{i,n}\delta_{l,m}                                      \eqno(2.2)
$$
with similar definition for $\overline S'(q,p)$. With the help of these
matrices we can write three more forms of (1.1).

 First, multiply both sides of $(1.1a)$ by
$S'(r,p)_{j_3,l_1}^{s_3,i_1}\overline S'(r,p)_{l_3,j_1}^{i_3,s_1}$
and sum over $i_1,j_1,i_3,j_3$. Then, using (2.2), we obtain
$$
\eqalign{
&\sum_{i_1,k_2,j_3}S(p,q)_{i_1,i_2}^{s_1,k_2}
S'(r,p)_{j_3,l_1}^{s_3,i_1}
S(q,r)_{k_2,l_3}^{j_2,j_3}\cr
=&R_{p,q,r}\sum_{j_1,k_2,i_3}
\overline S(q,r)_{i_2,i_3}^{k_2,s_3}
\overline S'(r,p)_{l_3,j_1}^{i_3,s_1}
\overline S(p,q)_{l_1,k_2}^{j_1,j_2}.\cr}           \eqno(2.3)
$$
Next, multiplying (2.3) by
$S'(r,q)_{i_3,j_2}^{l_3,s_2}\overline S'(r,q)_{s_3,l_2}^{j_3,i_2}$,
summing over $i_2,j_2,l_3,s_3$, and using again (2.2), we obtain another
 form of (1.1)
$$
\eqalign{
&\sum_{i_1,i_2,s_3}S(p,q)_{i_1,i_2}^{s_1,s_2}
S'(r,p)_{i_3,l_1}^{s_3,i_1}
\overline S'(r,q)_{s_3,l_2}^{j_3,i_2}\cr
 =&R_{p,q,r}\sum_{j_1,j_2,l_3}
S'(r,q)_{i_3,j_2}^{l_3,s_2}
\overline S'(r,p)_{l_3,j_1}^{j_3,s_1}
\overline S(p,q)_{l_1,l_2}^{j_1,j_2}.\cr}           \eqno(2.4)
$$
At last, multiplication of (2.3) by
$S'(q,p)_{l_2,s_1}^{i_2,j_1}\overline S'(q,p)_{j_2,i_1}^{s_2,l_1}$,
summation over $l_1,s_1,i_2,j_2$, and the use of (2.2) lead to
$$
\eqalign{
 &\sum_{l_1,j_2,j_3}\overline S'(q,p)_{l_2,i_1}^{s_2,l_1}
S'(r,p)_{j_3,l_1}^{s_3,j_1}
S(q,r)_{l_2,l_3}^{j_2,j_3}\cr
 =&R_{p,q,r}\sum_{s_1,i_2,i_3}
\overline S(q,r)_{i_2,i_3}^{s_2,s_3}
\overline S'(r,p)_{l_3,i_1}^{i_3,s_1}
S'(q,p)_{l_2,s_1}^{i_2,j_1}.\cr}           \eqno(2.5)
$$

Now introduce an $N^4$-by-$N^4$ matrix $\R(p,p';q,q')$ through a ``box''
construction known for solutions of the star-triangle relations [5]:
$$
\eqalign{
 &\langle \overline k,\overline l|{\cal R}(p,p';q,q')|\overline m,
\overline n\rangle\cr
 =&\sum_{i,i',j,j'}S(p,q)_{k_1,j}^{i,n_1}
S(q',p)_{n_2,i}^{j',m_1}
S(p',q')_{m_2,j'}^{i',l_2}
S'(q,p')_{l_1,i'}^{j,k_2},\cr}                \eqno(2.6)
$$
where $\overline k,\overline l,\overline m,\overline n$ are
two-component multi-indices, taking $N^2$ values:
$$
\overline k=(k_1,k_2),\quad k_1,k_2=0,\ldots,N-1,    \eqno(2.7)
$$
and similarly for $\overline l,\overline m,\overline n$.
We want to show that $\R(p,p';q,q')$ solves the following YBE:
$$
\eqalign{
 {\cal R}_{12}(p,p';q,q')&{\cal R}_{13}(p,p';r,r'){\cal
R}_{23}(q,q';r,r')\cr =&{\cal R}_{23}(q,q';r,r'){\cal
R}_{13}(p,p';r,r'){\cal R}_{12}(p,p';q,q').\cr}           \eqno(2.8)
$$
To do that, define auxiliary $N^2$-by-$N^2$ matrices $U_{i,j}(p;q,r)$,
$V_{i,j}(p;q,r)$ by
$$
\langle\overline k|U_{i,j}(p;q,r)|\overline l\rangle=
\sum_s S(p,q)_{i,k_1}^{s,l_1}S(r,p)_{l_2,s}^{k_2,j},       \eqno(2.9)
$$
$$
\langle\overline k|V_{i,j}(p;q,r)|\overline l\rangle=
\sum_s S'(q,p)_{k_1,s}^{l_1,i}S(p,r)_{j,l_2}^{s,k_2}.       \eqno(2.10)
$$
Using consequently eqs. (2.3), and three times (1.1), we easily come
to the following relation:
$$
\eqalign{
\sum_k&U_{i,k}(r;p,p')\otimes U_{k,j}(r;q,q')\R(p,p';q,q')\cr
 =&\rho_{p,p',q,q',r}\overline\R(p,p';q,q')\sum_k U_{k,j}(r;p,p')\otimes
                                                   U_{i,k}(r;q,q'),\cr}
                                                            \eqno(2.11a)
$$
where $\overline\R(p,p';q,q')$ being defined as in (2.6) with all $S$'s
replaced by $\overline S$'s, and
$$
\rho_{p,p',q,q',r}=R_{p',r,q}R_{r,p,q}R_{p',q',r}/R_{q',r,p}.  \eqno(2.11b)
$$
Similarly, using consequently eqs. (2.4),
 (2.5), (1.1), and (2.3), we obtain
$$
\eqalign{
\rho'_{p,p',q,q',r'}
\sum_k&V_{i,k}(r';p,p')\otimes V_{k,j}(r';q,q')\overline\R(p,p';q,q')\cr
 =&\R(p,p';q,q')\sum_k V_{k,j}(r';p,p')\otimes V_{i,k}(r';q,q'),\cr}
                                                            \eqno(2.12a)
$$
where
$$
\rho'_{p,p',q,q',r'}=R_{r',p',q'}R_{r',p,q}R_{r',q',p}/R_{r',p',q}.
                                                             \eqno(2.12b)
$$
As a consequense of (1.3) one obtains
$$
(\rho_{p,p',q,q',r}/\rho'_{p,p',q,q',r'})^{N^3}=1,    \eqno(2.13)
$$
so $\rho$'s coinside up to some $N^3$-th root of unity.
On the other hand, at the particular choice $p=p'=q=q'=r=r'$ they coinside
themselves,
so by continuity argument we conclude that
$$
\rho_{p,p',q,q',r}=\rho'_{p,p',q,q',r'}.                \eqno(2.14)
$$
Using this fact and
$$
\langle \overline i,\overline j|{\cal R}(p,p';q,q')|\overline k,
\overline l\rangle=
\langle\overline j|V_{i_2,k_2}(p';q,q')
U_{i_1,k_1}(p;q,q')|\overline l\rangle,                       \eqno(2.15)
$$
we see that YBE (2.8) holds as a consequence of (2.11) and (2.12).

To construct a commuting family of transfer matrices,
 fix some positive integer $L$ and introduce a two-layer transfer matrix
$T(p,p')$ by
$$
\langle\overline i_1,\ldots,\overline i_L|T(p,p')|
\overline j_1,\ldots,\overline j_L\rangle
 =\sum_{\overline k_1,\ldots,\overline k_L}\prod_{s=1}^L
\langle \overline k_s,\overline i_s|{\cal R}(p,p';q,q')|\overline k_{s+1},
\overline j_s\rangle                                  \eqno(2.16)
$$
with periodicity condition $\overline k_1=\overline k_{L+1}$. Using YBE
 (2.8) in the standard way [1], we obtain a commutativity condition
$$
T(p,p')T(q,q')=T(q,q')T(p,p').                          \eqno(2.17)
$$
So, the lattice model, corresponding to $R$-matrix (2.6) is integrable.
In the next section we write down solutions to (1.1) and thereby
realize explicitly constructions of the present section.

\beginsection{3. Some Solutions}

Here we present particular solutions to eqs.(1.1).

Denote
$$
\w=\exp(2\pi i/N).                                        \eqno(3.1)
$$
 First, following
[6], remind an explicit form of Boltzmann weights of the $sl(3)$ chiral
Potts model. Let $\Gamma$ be an algebraic curve, defined by the following
equations in ${\bf P}^5$:
$$
\Gamma:\quad\pmatrix{(h^+_i)^N\cr (h^-_i)^N\cr}=K_iK_j^{-1}
\pmatrix{(h^+_j)^N\cr (h^-_j)^N\cr},\quad i,j\in Z_3=\{0,1,2\}, \eqno(3.2)
$$
where $h^\pm_i$ are homogeneous coordinates in ${\bf P}^5$, and
$K_i$, moduli 2-by-2 matrices with one and the same determinant for all
$i$'s:
$$
 \det K_i=\det K_j,\quad i,j\in Z_3.                       \eqno(3.3)
$$ We impose additional constraints on moduli matrices. Let
$$
K_i=\pmatrix{a_i\quad b_i\cr c_i\quad d_i\cr},\quad i\in Z_3.\eqno(3.4)
$$
Introduce one more set of matrices:
$$
K'_i=\pmatrix{a_{i-1}\quad b_{i-1}\cr c_i\quad d_i\cr},\quad i\in Z_3,
                                                             \eqno(3.5)
$$
and demand for them the same conditions as (3.3):
$$
\det K'_i=\det K'_j,\quad i,j\in Z_3.                         \eqno(3.6)
$$
Besides, define an algebraic curve $\Gamma'$ by (3.2) with $K'_i$'s instead
of $K_i$'s. Obviously, $\Gamma$ and $\Gamma'$ are birationally isomorphic:
$$
\tau:\Gamma\to\Gamma'
$$
$$
\tau^*(h^+_i)=h^+_{i-1},\quad\tau^*(h^-_i)= h^-_i,\quad i\in Z_3.\eqno(3.7)
$$

 For $p,q\in\Gamma$ or $\Gamma'$ and
$$
\overline m=(m_1,m_2)\in (Z_N)^2:\quad m_1,m_2=0,\ldots,N-1 \pmod N
                                                           \eqno(3.8)
$$
define a function $\overline W_{p,q}(\overline m)$ by the following
 relations:
$$
 {\overline W_{p,q}(\overline m+\overline\delta_1)\over
\overline W_{p,q}(\overline m)}=
 {h^+_0(p)h^-_0(q)-h^+_0(q)h^-_0(p)\w^{-m_1}\over
h^+_1(p)h^-_1(q)-h^+_1(q)h^-_1(p)\w^{1+m_1-m_2}},              \eqno(3.9a)
$$
$$
 {\overline W_{p,q}(\overline m+\overline\delta_2)\over
\overline W_{p,q}(\overline m)}=
 {h^+_1(p)h^-_1(q)-h^+_1(q)h^-_1(p)\w^{m_1-m_2}\over
h^+_2(p)h^-_2(q)-h^+_2(q)h^-_2(p)\w^{1+m_2}},                  \eqno(3.9b)
$$
where $\overline\delta_1=(1,0)$ and $\overline\delta_2=(0,1)$, and
$$
\overline W_{p,q}(0)=1.                                    \eqno(3.9c)
$$
With these definitions, Boltzmann weights of the $sl(3)$ chiral Potts model
have the following form:
$$
\overline W_{p,q}(\overline m,\overline n)=
\w^{(m_2-n_2)(n_1-n_2)-(m_1-n_1)n_1}
\overline W_{p,q}(\overline m-\overline n).                 \eqno(3.10)
$$

Now we formulate the main result of the paper.\hfill\break
 {\bf Theorem}. {\it The following functions satisfy} (1.1):
$$
S(p,q)^{n_1,n_2}_{m_2,m_1}=
\lambda(\overline m)\overline W_{p,q}(\overline m,\overline n)/
\lambda(\overline n),                                       \eqno(3.11a)
$$
$$
\overline S(p,q)^{n_1,n_2}_{m_2,m_1}=
\lambda'(\overline m)\overline W_{\tau(p),\tau(q)}(\overline m,\overline n)/
\lambda'(\overline n),                                      \eqno(3.11b)
$$
 {\it where $p,q\in\Gamma$, and}
$$
\lambda(\overline m)=\w^{-2m_1m_2-m_2(m_2+N)/2},\quad
\lambda'(\overline m)=\lambda(\overline m)\w^{3m_1m_2}.
                  \eqno(3.12)
$$
An explicit form of the function $R_{p,q,r}$ and the proof of the theorem
is given in appendix A. Note, that, if all moduli matrices are
lower-triangular, then $\Gamma'=\Gamma$ and $\tau={\rm id}$. In this case,
and when $N=3$, we have $R_{p,q,r}=1$, $\overline S_{p,q}=S_{p,q}$. So,
the theorem gives a solution to the usual ``vertex'' YBE.

All the constructions of section 2 can be performed with solutions, given
by the theorem, only for the {\bf odd} $N$, since for even $N$ the
conditions (2.1) do not hold (see appendix B).

\beginsection{4. Summary}

In this paper we have shown that the generalized YBE
 (1.1), being in fact a ``vertex'' counterpart of the star-triangle equations
 (1.4),
admit non-trivial solutions, given by (3.11). The latter reduce to the
usual $R$-matrix, i. e. the solution of the ``vertex'' YBE, if the moduli
matrices (3.4) are lower-triangular, and the number of local states
$N=3$.

By the use of matrix elements of the $S$-matrix from (1.1) it
is possible to build integrable models on square
lattice with commuting two-layer transfer matrices, single-layer ones being
not necessarily so. More precisely, solutions of eqs.  (1.1) should
satisfy also conditions (2.1) (this is the case for solutions (3.11) only
 for odd number of local states).

The family of two-layer transfer matrices (2.16) is commutative provided
the ``box'' $R$-matrix (2.6) solves the YBE (2.8). The latter is a
consequence of the generalized YBE (1.1).

In conclusion note that from the physical point of view the lattice model,
constructed in this paper, is unsatisfactory, since there is no region in
the parameter space where the Botzmann weights are non-negative. The same
problem we have also in the $sl(n)$ chiral Potts model for $n\ge3$. Whether
one can find a similarity transformation, leading to a physical
lattice model, is an open question.

\beginsection{Acknowlegements}

We would like to thank L.D. Faddeev,
A.Yu. Alekseev, and E.K. Sklyanin for pointing out refs. [3,4]
and useful discussion of the results of this paper.

\beginsection{References}

\item{[1]}
R.J. Baxter, {\it ``Exactly Solved Models in Statistical Mechanics'',}
Academic Press, London, 1982.

\item{[2]}
 Yang, C.N.: Phys. Rev. Letters, {\bf19}, 1312 (1967)

\item{[3]}
Gervais, J.-L., Neveu, A.: Nucl. Phys. B {\bf238} (1984)\hfill\break
Alekseev, A.Yu., Faddeev, L.D.: Commun. Math. Phys. {\bf141}, 413 (1991)

\item{[4]}
 Drinfeld, V.G.: Algebra and Analysis, Vol. {\bf1}, No. 6, 114 (1989)
 (in Russian)

\item{[5]}
Baxter, R.J, Perk, J.H.H., Au-Yang, H.: Phys. Lett. {\bf A128}, 138 (1988)

\item{[6]}
Bazhanov, V.V., Kashaev, R.M., Mangazeev, V.V., Stroganov, Yu.G.:
Commun. Math. Phys. {\bf138}, 393 (1991)

\item{[7]} Bazhanov, V.V., Baxter, R.J.: {\it``New solvable lattice models in
three dimensions.''} ANU preprint MMR-5-92/SMS-15-92, Canberra, 1992.

\item{[8]}
R.M. Kashaev, V.V. Mangazeev and Yu.G. Stroganov,
{\it ``Star-Square and Tetrahedron Equations
in the Baxter-Bazhanov Model'',}
Preprint IHEP-92-135, Protvino, 1992, to appear in Intern. J. Mod. Phys. A.

\beginsection{Appendix A}

In this appendix we prove the theorem of section 3. First, following
[7] and [8], introduce more auxiliary objects.
 For any complex $x$ and integer $l$ define
$$
w(x|0)=1,\quad w(x|l)=\prod_{j=1}^l{1\over(1-x\w^j)}.       \eqno(A.1)
$$
This function has specific properties
$$
w(x|l+m)=w(x|l)w(x\w^l|m),                          \eqno(A.2a)
$$
$$
 w(x/\w|l)w(1/x|-l)=\w^{l(l+1)}(-x)^l.                \eqno(A.2b)
$$
With the aid of definition $(A.1)$ introduce
$$
 f(x,y|z)=\sum_{\s=0}^{N-1}{w(x|\s)\over w(y|\s)}z^\s,         \eqno(A.3)
$$
where complex parameters $x,y,z$ are constrained by the relation
$$
z^N={1-x^N\over1-y^N}  ,                                       \eqno(A.4)
$$
providing periodicity of the summand in $(A.1)$ on variable $\s$ with period
$N$. In [8] the following automorphisity property
of $f(x,y|z)$ is proved, see $(A.14)$ there:
$$
 {f(x\w^k,y\w^l|z\w^m)\over f(x,y|z)}={w(x/y\w|k-l)w(y|l)w(1/z|-m)\over
z^ly^m\w^{m(l+1)}w(x|k)w(x/yz\w|k-l-m)}.                        \eqno(A.5)
$$
Now we turn to the proof of (1.1).
To simplify formulae below, we will use the following notations:
$$
W_{pq}(m_1,m_2)\equiv\overline W_{p,q}(\overline m),\quad
W'_{pq}(-m_2,-m_1)\equiv\overline W_{\tau(p),\tau(q)}(\overline m).
                                                            \eqno(A.6)
$$
Substituting (3.11), (3.12) into $(1.1a)$ and cancelling common factors, we
obtain:
$$
\eqalign{
 &{\rm L.H.S.}=\sum_{k_1,k_2,k_3}\Phi(i_2)\Phi(k_1)\Phi(k_2)^2\Phi(k_3)^3
\w^{k_1k_2-k_1k_3-2k_2k_3}\cr
\times&
\w^{k_1(j_1+i_1-i_2+i_3)+k_2(j_2-j_3-i_1)-k_3(2j_1+j_2)+j_2j_3-j_1i_3-2i_1i_2}
\cr
\times&
W_{pq}(i_2-k_1,i_1-k_2)W'_{pr}(k_3-k_1,j_1-i_3)W_{qr}(k_3-j_2,k_2-j_3),\cr}
                                                          \eqno(A.7a)
$$
and
$$
 {\rm R.H.S.}=R_{p,q,r}\times({\rm L.H.S.}\quad{\rm with}\quad
i_s\leftrightarrow j_s, \quad{\rm and}\quad W\leftrightarrow W'),\eqno(A.7b)
$$
where
$$
\Phi(k)=\w^{k(k+N)/2}.                                     \eqno(A.8)
$$
Below we will transform explicitly only the L.H.S. up to factors,
symmetrical with respect to interchange
$i_s\leftrightarrow j_s, \quad{\rm and}\quad W\leftrightarrow W'$
, simultaneously assuming that
the R.H.S. is transformed in accordance with $(A.7b)$.

Let us shift indices $i_2,j_2,k_1,k_3$ by one and the same
quantity in $(A.7)$:
$$
i_2\to i_2+s,\quad j_2\to j_2+s,\quad k_1\to k_1+s,\quad k_3\to i_2+s.
                                                          \eqno(A.9)
$$
Up to symmetrical factors, the only change we have is
$$
 {\rm summand}\quad {\rm of}\quad
 {\rm L.H.S.}\sim\w^{s(i_2+k_3-k_1)}.                       \eqno(A.10)
$$
Multiplying now the both sides by $\w^{st}$ and summing over $s$, one
gets instead of $(A.10)$
$$
 {\rm summand\quad of\quad L.H.S.}\sim\delta(t+i_2+k_3-k_1),\eqno(A.11)
$$
so the summations over $k_1$ can be performed trivially. Changing
variables
$$
k_2\to k_2+j_3,\quad k_3\to k_3+j_2                        \eqno(A.12)
$$
and after that,
$$
i_2\to -i_2-t,\quad j_2\to -j_2-t,\quad
i_3\to -i_3+j_1,\quad j_3\to -j_3+i_1,                     \eqno(A.13)
$$
rewrite $(A.7)$ as follows
$$
\eqalign{
 {\rm L.H.S.}=W'_{pr}(i_2,i_3)\sum_{k_2,k_3}
 &W_{pq}(j_2-k_3,j_3-k_2)W_{qr}(k_3,k_2)\cr
\times&\w^{k_2^2+k_3^2-k_2k_3-k_2(i_2+j_3)+k_3(i_2-i_3-j_2+j_3)}.\cr}
                                                               \eqno(A.14)
$$
To proceed further, we need an explicit form of $W_{pq}(k,l)$ in terms of
 functions defined in $(A.1)$. Solving
recurrence relations (3.9), one can write
$$
W_{pq}(k,l)=w(x_{pq}|k-l)w(y_{pq}|l)w(z_{pq}|-k)u_{pq}^l/v_{pq}^k,
                                        \eqno(A.15)
$$
where parameters $x_{pq},\ldots,v_{pq}$ can be expressed in terms of
original variables $h^\pm_i(p),\quad h^\pm_i(q)$. Note only, that as a
consequence of (3.2), the parameters in $(A.15)$ satisfy the following
periodicity conditions:
$$
u_{pq}^N=(1-y_{pq}^N)/(1-x_{pq}^N),\quad
v_{pq}^N=(1-z_{pq}^N)/(1-x_{pq}^N).                \eqno(A.16)
$$
If we define the symbol $\overline{pq}$ as an abstract notation for a new
set of variables:
$$
x_{\overline{pq}}=1/\w x_{pq},\quad
y_{\overline{pq}}=1/\w y_{pq},\quad
z_{\overline{pq}}=1/\w z_{pq},                         \eqno(A.17a)
$$
$$
u_{\overline{pq}}=u_{pq}x_{pq}/y_{pq},\quad
v_{\overline{pq}}=v_{pq}x_{pq}/z_{pq},                 \eqno(A.17b)
$$
which satisfy relations $(A.16)$ as well,
then, applying $(A.2b)$ to $(A.15)$, we obtain
$$
W_{pq}(k,l)=\w^{-k^2-l^2+kl}/W_{\overline{pq}}(-k,-l).   \eqno(A.18)
$$
The corresponding expression for the $W'_{pq}(k,l)$ has the same form $(A.15)$
with the all parameters being primed.

Let us transform $W_{qr}(k_3,k_2)$ in $(A.14)$ with the aid of $(A.18)$.
Using of $(A.15)$ and dividing by $W_{pr}(j_2,j_3)W'_{pr}(i_2,i_3)$ we have
$$
 {\rm L.H.S.}={W_{pq}(j_2,j_3)\over W_{pr}(j_2,j_3)}
\sum_{k_1,k_2,k_3}\delta(k_1+k_2+k_3)
\prod_{s=1}^3\xi_s^{k_s}{w(x_s|k_s)\over w(\overline x_s|k_s)},
                                                          \eqno(A.19)
$$
where
$$
 \eqalign{x_1&=x_{pq}\w^{j_2-j_3},\cr
\overline x_1&=x_{\overline{qr}},\cr
        \xi_1&=\xi\w^{-i_2-j_3},\cr}\quad
 \eqalign{x_2&=y_{pq}\w^{j_3},\cr
\overline x_2&=y_{\overline{qr}},\cr
        \xi_2&=\xi u_{pq}/u_{\overline{qr}},\cr}\quad
 \eqalign{x_3&=z_{pq}\w^{-j_2},\cr
\overline x_3&=z_{\overline{qr}},\cr
        \xi_3&=\xi v_{pq}\w^{-i_3-j_2}/v_{\overline{qr}},\cr} \eqno(A.20a)
$$
and $\xi$ is any root of the equation
$$
\xi^N=(1-x_{pq}^N)/(1-x_{\overline{qr}}^N).  \eqno(A.20b)
$$ In $(A.19)$ we
have negated $k_2$, added one more summation over $k_1$ together with
delta-symbol $\delta(k_1+k_2+k_3)$, and distributed the remained factors
among $\xi_1,\xi_2,\xi_3$. To make the last step, remind the following
expression for the delta-symbol:
$$
N\delta(k)=\sum_{\sigma}\w^{k\sigma}.                 \eqno(A.21)
$$ Substituting it
into $(A.19)$ and using definition $(A.3)$ together with $(A.4)$, we
 finally obtain
$$
 {\rm L.H.S.}={W_{pq}(j_2,j_3)\over W_{pr}(j_2,j_3)}
\prod_{s=1}^3f(x_s,\overline x_s|\xi_s)
\sum_{\sigma}\prod_{s=1}^3x_s^{-\sigma}
 {w(\overline x_s\xi_s/x_s|k_s)\over w(\xi_s/\w|k_s)}.       \eqno(A.22)
$$
Restoring the R.H.S. through $(A.7b)$ and expressing all parameters in terms
of original ones, we convince that L.H.S.=R.H.S., the $R$-factor being given
by
$$
R_{p,q,r}=\prod_{s=1}^3f(x_s,\overline x_s|\xi_s)
/f(x'_s,\overline x'_s|\xi'_s)\Biggr|_{i_2=i_3=j_2=j_3=0}.
                                                              \eqno(A.23)
$$

\beginsection{Appendix B}

Here we examine validity of conditions (2.1) for solutions (3.11).

Consider the linear equations
$$
\sum_{m_1,n_1}S(p,q)^{n_1,n_2}_{m_2+n_2,m_1+n_1}\w^{m_1n_1}\Psi_{m_1,n_1}
 =0                                                               \eqno(B.1)
$$
on $N^2$ unknowns $\Psi_{m_1,n_1}$. The number of linearly independent
solutions of eqs. $(B.1)$ is equal to $N^2-{\rm{\bf rank}}S(p,q)^{t_2}$,
so the first condition in (2.1) is equivalent to absence of non-zero
solutions. Substituting $(3.11a)$, and omitting non-zero factors, we rewrite
$(B.1)$ as
$$
\sum_{m_1,n_1}\w^{-(2m_2+2n_2)m_1-m_2n_1}\overline
W_{p,q}(\overline m)\Psi_{m_1,n_1}=0.                            \eqno(B.2)
$$
To solve these relations introduce a new set of $N^2$ unknowns:
$$
\widetilde{\Psi}_{m_2,n_2}=\sum_{m_1,n_1}\w^{-(2m_2+n_2)m_1-m_2n_1}\overline
W_{p,q}(\overline m)\Psi_{m_1,n_1},                          \eqno(B.3)
$$
the inverse transformation being given by
$$
\Psi_{m_1,n_1}=\sum_{m_2,n_2}\w^{(2m_2+n_2)m_1+m_2n_1}{1\over N^2\overline
W_{p,q}(\overline m)}\widetilde{\Psi}_{m_2,n_2}.           \eqno(B.4)
$$
Comparing $(B.2)$ and $(B.3)$, we immediately conclude, that $(B.2)$ have a
very simple form in the new variables:
$$
\widetilde{\Psi}_{m_2,2n_2}=0.                               \eqno(B.5)
$$
 For odd $N$ variable $2n_2$ runs over all $N$ values $0,1,\ldots,N-1$, while
 for even $N$, only $N/2$ values $0,2,\ldots,N-2$. Thus, we have proved that
$$
{\rm{\bf rank}}S(p,q)^{t_2}=\cases{N^2,&if $N=1\pmod2$;\cr N^2/2,&if
                                 $N=0\pmod2$.\cr}            \eqno(B.6)
$$
Analogous consideration leads to the same result for
$\overline S(p,q)^{t_2}$.  \bye